\documentstyle[12pt,amstex]{article} 
  \newtheorem{df}{Definition}[section]
  
  \newtheorem{lem}[df]{Lemma}
  \newtheorem{rmk}[df]{Remark}
  \newtheorem{prop}[df]{Proposition}
  \newtheorem{cor}[df]{Corollary}
  
  \makeatletter
   
   \@addtoreset{equation}{section}
  \makeatother

\begin{document}
 \title{ Submodels of Nonlinear Grassmann \\
         Sigma Models in Any Dimension \\
         and \\
         Conserved Currents, Exact Solutions }
 \author{ 
  Kazuyuki FUJII,\thanks{Department of Mathematics, 
  Yokohama City University, 
  Yokohama 236, 
  Japan, \endgraf 
  {\it E-mail address}: fujii{\char'100}yokohama-cu.ac.jp} \ 
  Yasushi HOMMA\thanks{Department of Mathematics, 
  Waseda University, 
  Tokyo 169, 
  Japan, \endgraf 
  {\it E-mail address}: 696m5121{\char'100}mn.waseda.ac.jp} \ and 
  Tatsuo SUZUKI\thanks{Department of Mathematics, 
  Waseda University, 
  Tokyo 169, 
  Japan, \endgraf 
  {\it E-mail address}: 695m5050{\char'100}mn.waseda.ac.jp}}
 \date{}
 \maketitle
\begin{abstract}
In the preceding paper(hep-th/9806084), we constructed submodels of nonlinear Grassmann sigma models in any dimension and, moreover, an infinite number of conserved currents and a wide class of exact solutions.

In this paper, we first construct almost all conserved currents for the submodels and all ones for the one of ${\bf C}P^1$-model. We next review the Smirnov and Sobolev construction for the equations of ${\bf C}P^1$-submodel and extend the equations, the S-S construction and conserved currents to the higher order ones.
\end{abstract}
 \section{Introduction}
This paper is a sequel to the proceding paper \cite{FHS}. In \cite{FHS}, for nonlinear Grassmann sigma models in any dimensions, we defined their submodels with both global and local form:
$$
 [P, \partial^{\mu}\partial_{\mu}P]=0 \quad \mbox{and} \quad
 [P \otimes P, \partial^{\mu}\partial_{\mu}(P \otimes P)]=0
$$
$$ \hspace{5cm} \mbox{for} \quad
 P:M^{1+n} \longrightarrow G_{j,N}({\bf C}),
$$
globally and
$$
 \partial^{\mu}\partial_{\mu}Z=0 \quad \mbox{and} \quad 
 \partial^{\mu}Z \otimes \partial_{\mu}Z=0
$$
$$ \hspace{5cm} \mbox{for} \quad
 Z:M^{1+n} \longrightarrow M(N-j,j;{\bf C})
$$
locally.

For the submodels, we constructed an infinite number of conserved currents and a wide class of exact solutions. This is a generalization of ones in previous papers \cite{AFG}, \cite{FS1}, \cite{FS2}, \cite{GMG}. But, we didn't determine all conserved currents or all exact solutions.

In the former part of this paper, we construct almost all conserved currents for our submodels and all ones for the submodel of ${\bf C}P^1$-model (${\bf C}P^1$-submodel for simplicity).

By the way, we knew by \cite{BY} that the construction of exact solution for the equation of ${\bf C}P^1$-submodel;
$$
 \partial^{\mu}\partial_{\mu}u=0 \quad \mbox{and} \quad 
 \partial^{\mu}u\partial_{\mu}u=0
$$
$$ \hspace{5cm} \mbox{for} \quad
 u:M^{1+n} \longrightarrow {\bf C}
$$
has been known by Smirmov and Sobolev \cite{SS1}, \cite{SS2}.

We review their construction briefly and give a key lemma to construct explicit (not abstruct) solutions. 

On the other hand, it is easy for us to generalize the S-S construction itself. But what kind of equations fit to the generalized one? 

In the latter part of this paper, we give such equations which include higher order derivatives. We may be able to construct a new theory including higher order derivatives.

 \section{Submodels of Nonlinear Grassmann Sigma Models}
We review \cite{FHS} within our necessity. Let $M(m,n;{\bf C})$ be the set of $m \times n$ matrices over $\bf C$ and $M(n;{\bf C})$ be $M(n,n;{\bf C})$ for simplicity. For a pair $(j,N)$ with $1 \le j \le N-1$, we set $I$, $O$ as a unit matrix, a zero matrix in $M(j;{\bf C})$ and  $I'$, $O'$ as ones in $M(N-j;{\bf C})$ respectively. 

We define a Grassmann manifold for the pair $(j,N)$ above as
\begin{equation}
 G_{j,N}({\bf C}) \equiv \{ P \in M(N;{\bf C})| P^2=P, P^{\dag}=P, \mbox{tr}P=j \}.
\end{equation}
Then it is well-known that
\begin{eqnarray}
 G_{j,N}({\bf C}) 
 &=&
 \left\{ U
          \left(
           \begin{array}{cc}
               I &    \\
                 & O' \\
            \end{array}
           \right)
            U^{\dag}|U \in U(N) \right\}  \label{eqn:1-2}\\
 &\cong& \displaystyle{\frac{U(N)}{U(j) \times U(N-j)}}.\label{eqn:1-3}
\end{eqnarray}
Next, we introduce a local chart for $G_{j,N}({\bf C})$. For $Z \in M(N-j,j;{\bf C})$ a neighborhood of 
         $\left(
           \begin{array}{cc}
               I &    \\
                 & O' \\
            \end{array}
           \right) $
is expressed as 
\begin{equation}
 P_0(Z)=
          \left(
           \begin{array}{cc}
               I & -Z^{\dag}  \\
               Z & I' \\
            \end{array}
           \right)  
          \left(
           \begin{array}{cc}
               I &    \\
                 & O' \\
            \end{array}
           \right)  
          \left(
           \begin{array}{cc}
               I & -Z^{\dag} \\
               Z & I' \\
            \end{array}
           \right)^{-1} ,
\end{equation}
so we have
\begin{equation}
 P(Z)=U P_0(Z) U^{\dag} \quad \mbox{for some} \ U \in U(N).
\end{equation}

Let $M^{1+n}$ be a $(1+n)$-dimensional Minkowski space $(n \in {\bf N})$ with a metric $\eta = (\eta_{\mu \nu}) = \mbox{diag}(1,-1,\cdots ,-1)$. Here we consider a nonlinear Grassmann sigma model on $M^{1+n}$. Let the pair $(j,N)$ be fixed. The action is 
\begin{equation}
 \mbox{$\cal{A}$}(P) \equiv \frac12 \int d^{1+n} x \ 
                             \mbox{tr}\partial_{\mu}P\partial^{\mu}P
 \label{eqn:1-6}
\end{equation}
$$ \hspace{5cm} \mbox{for} \quad
 P:M^{1+n} \longrightarrow G_{j,N}({\bf C}).
$$
Its equations of motion read 
\begin{equation}
 [P, \Box P] \equiv [P, \partial_{\mu}\partial^{\mu}P]=0.
\end{equation}
From this, we see that $[P, \partial_{\mu}P]$ are conserved currents (Noether currents).

Next, in \cite{FHS}, we defined a submodel of this as simultaneous equations
$$
 [P, \Box P]=0 \quad \mbox{and} \quad
 [P \otimes P, \Box (P \otimes P)]=0
$$
or transforming them to be
\begin{equation}
 [P, \Box P]=0 \quad \mbox{and} \quad
 [P,\partial_{\mu}P] \otimes \partial^{\mu}P 
 +\partial^{\mu}P \otimes [P,\partial_{\mu}P]=0.
 \label{eqn:1-8}
\end{equation}
\begin{rmk}
If $P$ satisfies (\ref{eqn:1-8}), then we find that $P \otimes P$ is a solution of $G_{j^2,N^2}$-model and, moreover, $\underbrace{P \otimes \cdots \otimes P}_k$ is the one of $G_{j^k,N^k}$-model.
\end{rmk}
\begin{rmk}
In fact, we can show that (\ref{eqn:1-8}) is equivalent to only one equation
$$
 [P \otimes P, \Box (P \otimes P)]=0.
$$
\end{rmk}
This is a global definition of our submodel. $P$ in (\ref{eqn:1-6}) is locally expressed as 
\begin{equation}
 P(Z)=U
         \left(
           \begin{array}{cc}
               I & -Z^{\dag}  \\
               Z & I' \\
            \end{array}
         \right)  
         \left(
           \begin{array}{cc}
               I &     \\
                 &  O' \\
            \end{array}
         \right) 
         \left(
           \begin{array}{cc}
               I & -Z^{\dag}  \\
               Z & I' \\
            \end{array}
         \right)^{-1}  U^{\dag}
\end{equation}
$$ \hspace{5cm} \mbox{for} \quad
 Z:M^{1+n} \longrightarrow M(N-j,j;{\bf C}),
$$
so that (\ref{eqn:1-8}) are written as
\begin{equation}
 \partial^{\mu}\partial_{\mu}Z=0 \quad \mbox{and} \quad 
 \partial^{\mu}Z \otimes \partial_{\mu}Z=0.
 \label{eqn:1-10}
\end{equation}
This is a local definition of our submodel. In the following, we write
\begin{equation}
 Z=(z_1,\cdots,z_M)^t, \quad M=(N-j)j
\end{equation}
as a vector in ${\bf C}^M$ for simplicity. Now, we again write down our equations of the submodel:
\begin{equation}
 \partial^{\mu}\partial_{\mu}Z=0 \quad \mbox{and} \quad 
 \partial^{\mu}Z \otimes \partial_{\mu}Z=0.
\end{equation}
$$ \hspace{5cm} \mbox{for} \quad
 Z:M^{1+n} \longrightarrow {\bf C}^M
$$
or in each component
\begin{equation}
 \partial^{\mu}\partial_{\mu}z_i=0 \quad \mbox{and} \quad 
 \partial^{\mu}z_i \partial_{\mu}z_j=0
 \label{eqn:1-13}
\end{equation}
for $1 \le i,j \le M$.

 \section{Conserved Currents}
In this section, we seek for all conserved currents for our submodels discussed in the previous section. First, we consider a one form on ${\bf C}^M$;
\begin{equation}
 \sum_{j=1}^M \{f_j (Z,\bar{Z})dz_j + g_j (Z,\bar{Z})d\bar{z}_j \}
\end{equation}
and pull back this on $M^{1+n}$ by $Z$;
$$
 Z:M^{1+n} \longrightarrow {\bf C}^M \quad Z=(z_j).
$$
Then, we look for conditions which make the pull-back form conserved currents;
\begin{equation}
 \sum_{j=1}^M \partial^{\mu}
  \{f_j (Z,\bar{Z})\partial_{\mu}z_j 
  + g_j (Z,\bar{Z})\partial_{\mu}\bar{z}_j \} =0.
\end{equation}
Making use of (\ref{eqn:1-13}), we have
\begin{prop}
if the equations
\begin{equation}
 \frac{\partial f_j}{\partial \bar{z}_k}
 +\frac{\partial g_k}{\partial z_j}=0
 \label{eqn:2-3}
\end{equation}
for $1 \le j,k \le M$ hold, then
\begin{equation}
 \sum_{j=1}^M 
  \{f_j (Z,\bar{Z})\partial_{\mu}z_j 
  + g_j (Z,\bar{Z})\partial_{\mu}\bar{z}_j \} .
\end{equation}
is conserved currents.
\end{prop}
Is the condition (\ref{eqn:2-3}) easy to solve? 
For example, we set
\begin{equation}
 f_j (Z,\bar{Z})=\frac{\partial f}{\partial z_j}, \quad
 g_j (Z,\bar{Z})=-\frac{\partial f}{\partial \bar{z}_j}
\end{equation}
for any function $f$ in $\mbox{C}^2$-class, then we can easily check (\ref{eqn:2-3}). Namely, we obtain conserved currents
\begin{equation}
 \sum_{j=1}^M 
  \left( \frac{\partial f}{\partial z_j}\partial_{\mu}z_j 
   -\frac{\partial f}{\partial \bar{z}_j}\partial_{\mu}\bar{z}_j \right) 
\end{equation}
parametrized by all $\mbox{C}^2$-class functions $f$. In particular, the situation becomes simpler in the case of ${\bf C}P^1$-submodel.
\begin{cor}
All conserved currents of the ${\bf C}P^1$-submodel are given by
\begin{equation}
  \frac{\partial f}{\partial z}\partial_{\mu}z 
   -\frac{\partial f}{\partial \bar{z}}\partial_{\mu}\bar{z} 
 \label{eqn:2-7}
\end{equation}
where $f=f(z,\bar{z})$ is any function in $\mbox{C}^2$-class.
\end{cor}
This corollary is proved by the Poincare lemma.

We have a bit complaint to this result. We obtained all conserved currents in terms of local form (\ref{eqn:1-10}). We don't know how to write down all of them in terms of global form (\ref{eqn:1-8}).

 \section{Exact Solutions $\cdots $ Smirnov and Sobolev Construction}
The submodel of ${\bf C}P^1$-model is defined by the equations
\begin{equation}
 \partial^{\mu}\partial_{\mu}u=0 \quad \mbox{and} \quad 
 \partial^{\mu}u\partial_{\mu}u=0
 \label{eqn:3-1}
\end{equation}
$$ \hspace{5cm} \mbox{for} \quad
 u:M^{1+n} \longrightarrow {\bf C}.
$$
For the equations, a wide class of explicit solutions have been constructed by Smirnov and Sobolev (S-S in the following) \cite{SS1}, \cite{SS2}. Let us make a short review according to \cite{BY} since we could not get old papers \cite{SS1}, \cite{SS2}. 

First, let $a_{\mu}(u), b(u)$ be known functions and we consider an equation
\begin{equation}
 0=\delta \equiv \sum_{\mu =0}^n a_{\mu}(u)x_{\mu}-b(u)
 \label{eqn:3-2}
\end{equation}
with the constraint
\begin{equation}
 a_0(u)^2-\sum_{j=1}^n a_j(u)^2=0.
 \label{eqn:3-3}
\end{equation}
Then, differentiating (\ref{eqn:3-2}), we have easily
\begin{equation}
 \partial_{\mu}u=-\frac{a_{\mu}}{\delta'}, \qquad \mbox{where} \quad
  {}'=\frac{\partial}{\partial u},
\end{equation}
\begin{equation}
 \partial_{\mu}^2 u=\frac{1}{{\delta'}^2}
  (2a_{\mu}a_{\mu}'-\frac{\delta''}{\delta'}a_{\mu}^2).
\end{equation}
Remarking (\ref{eqn:3-3}), we obtain (\ref{eqn:3-1}). Next, we solve (\ref{eqn:3-2}) making use of the inverse function theorem to be 
\begin{equation}
 u=u(x_0,x_1,\cdots,x_n).
 \label{eqn:3-6}
\end{equation}
For example, if $a_{\mu}(u)$ is a constant;
\begin{equation}
 a_0 x_0+\sum_{j=1}^n a_j x_j-b(u)=0
\end{equation}
with
\begin{equation}
 a_0^2-\sum_{j=1}^n a_j^2=0
\end{equation}
and $b$ has its inverse (setting $f$), then we have the well-known solutions
\begin{equation}
 u=f(a_0 x_0+\sum_{j=1}^n a_j x_j).
\end{equation}
This seems to be a main story of \cite{SS1}, \cite{SS2}. By the way, how do we solve (\ref{eqn:3-2}) as (\ref{eqn:3-6}) in an explicit manner? This is not so easy problem. For that we assume that $u$ is real analytic with respect to $(x_0,x_1,\cdots,x_n)$ (we choose $a_{\mu}(u), b(u)$ to become so).

Since $u$ is Taylor expansible, $u$ is completely determined if, for example, all coefficients at the origin;
$$
 \partial_{\mu_1} \cdots \partial_{\mu_k}u(0,0,\cdots,0), 
  \quad \quad k \in {\bf N}
$$
are determined. Therefore, we have only to determine $\partial_{\mu_1} \cdots \partial_{\mu_k}u$. Now let us define an operator
\begin{equation}
 X \equiv -\frac{\partial}{\partial u} \circ \frac{1}{\delta'}.
\end{equation}
Then
\begin{lem}
 we have
\begin{equation}
 \partial_{\mu_1} \cdots \partial_{\mu_k} u =
  -\frac{1}{\delta'} X^{k-1}(a_{\mu_1} \cdots a_{\mu_k}),
\end{equation}
where 
$$
\delta' = \sum_{\mu =0}^n a_{\mu}(u)'x_{\mu}-b(u)'.
$$
 \label{lem:3-1}
\end{lem}
The proof that is given by the mathematical induction is not so easy. Moreover,
\begin{prop}
 we have
\begin{equation}
 \partial_{\mu_1} \cdots \partial_{\mu_k} f(u) =
  -\frac{1}{\delta'} X^{k-1}
  (\frac{\partial f}{\partial u}a_{\mu_1} \cdots a_{\mu_k}),
\end{equation}
where 
$$
f:{\bf C} \longrightarrow {\bf C}:\mbox{holomorphic with respect to} \ u.
$$
 \label{prop:3-2}
\end{prop}
This formula will become useful in the next section when $f(u)=u^j$ for $1 \le j \le p$.

For the general equations
$$
 \partial^{\mu}\partial_{\mu}u_i=0 \quad \mbox{and} \quad 
 \partial^{\mu}u_i \partial_{\mu}u_j=0
  \qquad \mbox{for} \quad
 1 \le i,j \le M,
$$
we will discuss in an another paper.

 \section{Extension of Smirnov and Sobolev Construction}
In this section, we extend the S-S construction more generally. The S-S construction is 
$$
 0=\delta \equiv \sum_{\mu =0}^n a_{\mu}(u)x_{\mu}-b(u)
$$
with the constraint
$$
 a_0(u)^2-\sum_{j=1}^n a_j(u)^2=0.
$$
One can easily guess its extended form:
\begin{equation}
 0=\delta \equiv \sum_{\mu =0}^n a_{\mu}(u)x_{\mu}-b(u)
 \label{eqn:4-1}
\end{equation}
with the constraint ($p \geq 2$)
\begin{equation}
 a_0(u)^p-\sum_{j=1}^n a_j(u)^p=0.
 \label{eqn:4-2}
\end{equation}
On the other hand, the equations of S-S construction are
$$
\Box_2 u \equiv 
  \left(
   \frac{\partial^2}{\partial x_0^2}
   -\sum_{j=1}^n \frac{\partial^2}{\partial x_j^2}
  \right) u=0
\quad \mbox{and} \quad
 \partial^{\mu}u \partial_{\mu}u=0.
$$
Then, it is easy to see that these equations are equivalent to
\begin{equation}
\Box_2 u=0 \quad \mbox{and} \quad \Box_2 (u^2)=0. 
\end{equation}
What are equations satisfied by the extended S-S construction (\ref{eqn:4-1}), (\ref{eqn:4-2})? After some consideration, we reach the following equations.
\begin{df}
\begin{equation}
\Box_p (u^k) \equiv 
  \left(
   \frac{\partial^p}{\partial x_0^p}
   -\sum_{j=1}^n \frac{\partial^p}{\partial x_j^p}
  \right) (u^k)=0 \quad \mbox{for} \quad 1 \le k \le p .
 \label{eqn:4-4}
\end{equation}
\end{df}
Then, it is easy to check the next corollary by using proposition \ref{prop:3-2}.
\begin{cor}
Our extended S-S construction (\ref{eqn:4-1}), (\ref{eqn:4-2}) satisfies (\ref{eqn:4-4}).
\end{cor}
That is, our equations (\ref{eqn:4-4}) are ``integrable" equations. 
\begin{rmk}
By proposition \ref{prop:3-2}, we can show directly
$$
\Box_p f(u)=0,
$$
where 
$$
f:{\bf C} \longrightarrow {\bf C}:\mbox{holomorphic with respect to} \ u.
$$
But, in fact, if $u$ is a solution of (\ref{eqn:4-4}), then we have
$$
\Box_p f(u)=0.
$$
\end{rmk}
It is very interesting that the natural extension (\ref{eqn:4-1}), (\ref{eqn:4-2}) of S-S construction satisfies our new higher order equations (\ref{eqn:4-4}). We believe that this includes a deep fact. We need a further study.

 \section{Conserved Currents for New Higher Order Equations}
In this section, we construct conserved currents for the equations (\ref{eqn:4-4}).

Let $F_n$ be Bell polynomials (see Appendix in detail). We set $F_{n,\, \mu}$ as
\begin{eqnarray}
F_{n,\, \mu} & \equiv &
                :F_n(\partial_{\mu}u \frac{\partial}{\partial u},
                  \partial_{\mu}^2 u \frac{\partial}{\partial u},
                   \cdots ,
                  \partial_{\mu}^n u \frac{\partial}{\partial u}):
                  \nonumber \\
&=& \hspace{-8mm}
    \sum_{{\scriptstyle k_1+2k_2+ \cdots +nk_n=n}\atop
          {\scriptstyle k_1 \geq 0,\cdots ,k_n \geq 0}}
         \frac{n!}{k_1! \cdots k_n!} 
         \left(
          \frac{\partial_{\mu}u}{1!}
         \right)^{k_1}
         \left(
          \frac{\partial_{\mu}^2 u}{2!}
         \right)^{k_2}
            \cdots
         \left(
          \frac{\partial_{\mu}^n u}{n!}
         \right)^{k_n}
         \left(
          \frac{\partial }{\partial u}
         \right)^{k_1+k_2+ \cdots +k_n} \hspace{-20mm}, 
         \nonumber \\
  && 
\end{eqnarray}
and $\bar{F}_{n,\, \mu}$ its complex conjugate of $F_{n,\, \mu}$, where $: \ :$  means the normal ordering.
Then we have
\begin{prop}for $p \geq 2$,
\begin{equation}
V_{p,\, \mu}(f) \equiv 
  \sum_{k=0}^{p-1}(-1)^k :F_{p-1-k,\, \mu}\bar{F}_{k,\, \mu}:(f)
 \label{eqn:5-1}
\end{equation}
are conserved currents for the equations (\ref{eqn:4-4}), 
where $f=f(u,\bar{u})$ is any function in $C^p$-class. 
\end{prop}
For example, 
\begin{eqnarray}
 V_{2,\, \mu}(f) &=& F_{1,\, \mu}(f)-\bar{F}_{1,\, \mu}(f) \nonumber \\
   &=& \partial_{\mu}u \frac{\partial f}{\partial u}
      -\partial_{\mu}\bar{u} \frac{\partial f}{\partial \bar{u}},
      \hspace{75mm}
\end{eqnarray}
\begin{eqnarray}
 V_{3,\, \mu}(f) &=& F_{2,\, \mu}(f)
                    -:F_{1,\, \mu}\bar{F}_{1,\, \mu}:(f)
                    +\bar{F}_{2,\, \mu}(f) \nonumber \\
   &=& \partial_{\mu}^2 u \frac{\partial f}{\partial u}
      +(\partial_{\mu}u)^2 \frac{\partial^2 f}{\partial u^2}
      -\partial_{\mu}u \partial_{\mu}\bar{u} 
        \frac{\partial^2 f}{\partial u \partial \bar{u}}
      +\partial_{\mu}^2 \bar{u} \frac{\partial f}{\partial \bar{u}}
      +(\partial_{\mu}\bar{u})^2 \frac{\partial^2 f}{\partial \bar{u}^2}.
      \nonumber \\
   && 
\end{eqnarray}
Hence, (\ref{eqn:5-1}) is a generalization of (\ref{eqn:2-7}).

 \section{Discussion}
We in this paper constructed almost all conserved currents for the submodels of nonlinear Grassmann sigma models in any dimension. In particular, in the case of ${\bf C}P^1$-submodel, we determined all ones. 

We also in this case reviewed the S-S construction and gave the explicit form to the solutions. Moreover, we found new higher order equations to satisfy the extended S-S construction and constructed the conserved currents. 

A speculation from this work is now in order. A Dirac operator $D \hspace{-8pt}/ \ $ is defined by the square root of d'Alembertian $\Box_2$ ;
\begin{equation}
 D \hspace{-8pt}/ \, {}^2 \ = \Box_2 \qquad \mbox{on} \ M^{1+n}.
\end{equation}
In the following, we write $D \hspace{-8pt}/ \, {}_2 \ =D \hspace{-8pt}/ \ $ to emphasize $\Box_2$. From the extended S-S construction, we wish to define a $p$-th root of $\Box_p$ ;
\begin{equation}
 (D \hspace{-8pt}/ \, {}_p )^p = \Box_p \qquad \mbox{on} \ M^{1+n}.
\end{equation}
In fact, we can construct $D \hspace{-8pt}/ \, {}_p $ as
\begin{equation}
 D \hspace{-8pt}/ \, {}_p \equiv 
  \sum_{\mu =0}^n \Gamma_{\mu} \otimes \frac{\partial}{\partial x_{\mu}},
\end{equation}
and
\begin{equation}
 \Gamma_{\mu}\Gamma_{\nu}=e^{i\frac{2\pi}{p}}\Gamma_{\nu}\Gamma_{\mu} \ 
 (\mu < \nu ), \ \Gamma_0^p =1 \ 
 \mbox{and} \ \Gamma_j^p =-1 \ (1 \le j \le n),
\end{equation}
see \cite{R} and its references. From the construction of $\{ \Gamma_{\mu} \} $, $\Gamma_{\mu}$ is not hermitian, so that $D \hspace{-8pt}/ \, {}_p$ is not hermitian operator. Therefore, its physical meaning of $D \hspace{-8pt}/ \, {}_p$ is certainly not clear, but it may be worth studying in the mathematical point of view ($q$-deformation theory, etc). To reveal a deep meaning, it needs further consideration.
 \section*{Acknowledgements}
Kazuyuki Fujii was partially supported by Grant-in-Aid for Scientific Research (C) No. 10640210. KF is very grateful to Prof. Akira Asada for useful suggestions. Yasushi Homma and Tatsuo Suzuki are very grateful to Tosiaki Kori for valuable discussions. 

\appendix
\section{Appendix}
For variables $g_1 z, \cdots ,g_n z$, polynomials
\begin{equation}
F_n(g_1 z, \cdots ,g_n z) = \hspace{-5mm}
    \sum_{{\scriptstyle k_1+2k_2+ \cdots +nk_n=n}\atop
          {\scriptstyle k_1 \geq 0,\cdots ,k_n \geq 0}}
         \frac{n!}{k_1! \cdots k_n!} 
         \left(
          \frac{g_1 z}{1!}
         \right)^{k_1}
         \left(
          \frac{g_2 z}{2!}
         \right)^{k_2}
            \cdots
         \left(
          \frac{g_n z}{n!}
         \right)^{k_n}
\end{equation}
are called Bell polynomials. For example, 
$$
F_1=g_1 z, \qquad F_2=g_2 z + g_1^2 z^2.
$$
These polynomials are used in the differential calculations of composite functions. \cite{Ri}

Notice that we have a following recursion formula;
\begin{equation}
F_{n+1}=g_1 z F_n + \sum_{r=1}^n g_{r+1} \frac{\partial F_n}{\partial g_r}.
\end{equation}

  
\end{document}